\documentclass[acmsmall,screen,nonacm]{acmart}

\setcopyright{none}
\acmDOI{}
\acmISBN{}

\usepackage{amsmath}
\usepackage{wrapfig}

\usepackage{algorithm}
\usepackage{algpseudocode}

\usepackage{graphicx}
\usepackage{textcomp}
\usepackage{url}

\usepackage[table]{xcolor}

\usepackage{xspace}
\newcommand{\approach}{VisStateCoder\xspace}

\usepackage{booktabs}
\usepackage{multirow}
\usepackage{tabularx}
\usepackage{array}
\usepackage{siunitx}

\usepackage{tikz}

\usepackage{enumitem}
\setlist[itemize]{
    labelindent=2pt,
    leftmargin=*,
    labelsep=0.5em
}

\usepackage{pifont}

\definecolor{myitem}{HTML}{A64D79}

\usepackage[most]{tcolorbox}

\newtcolorbox{answerbox}{
  enhanced,
  colback=gray!8,
  colframe=gray!70!black,
  boxrule=1.8pt,
  arc=4pt,
  boxsep=0pt,
  left=1.4em,
  right=1.4em,
  top=0.9em,
  bottom=0.9em,
  width=\linewidth,
  before skip=0.8em,
  after skip=0.8em,
  fontupper=\small
}

\def\BibTeX{%
  {\rm B\kern-.05em%
  {\sc i\kern-.025em b}\kern-.08em%
  T\kern-.1667em%
  \lower.7ex\hbox{E}\kern-.125emX}%
}

\begin{document}

% =========================================================
% Title
% =========================================================

\title{State-Aware Generation of Deployable Websites
from Visual Prototypes and Requirements}

% =========================================================
% Authors
% =========================================================

% Equal first author
\author{Ruwei Pan}
\authornote{Both authors contributed equally to this work.}
\email{panruwei666@gmail.com}
\affiliation{%
  \institution{Peking University}
  \country{China}
}

% Equal first author
\author{Junlei Shen}
\authornotemark[1]
\email{liuy92246@gmail.com}
\affiliation{%
  \institution{Chongqing University}
  \country{China}
}

\author{Linhao Wu}
\email{wulinhao26@stu.pku.edu.cn}
\affiliation{%
  \institution{Peking University}
  \country{China}
}

\author{Zixiong Yang}
\email{zixiongyang02@gmail.com}
\affiliation{%
  \institution{Peking University}
  \country{China}
}

\author{Qisheng Zhang}
\email{20241025@stu.cqu.edu.cn}
\affiliation{%
  \institution{Chongqing University}
  \country{China}
}

\author{Yifan Li}
\email{liyifancqu@163.com}
\affiliation{%
  \institution{University of Science and Technology of China}
  \country{China}
}

\author{Tianyu Ma}
\email{20230574@stu.cqu.edu.cn}
\affiliation{%
  \institution{Chongqing University}
  \country{China}
}

\author{Yakun Zhang}
\email{zhangyk@hit.edu.cn}
\affiliation{%
  \institution{Harbin Institute of Technology (Shenzhen)}
  \country{China}
}

\author{Lu Zhang}
\email{zhanglucs@pku.edu.cn}
\affiliation{%
  \institution{Peking University}
  \country{China}
}

% Corresponding author
\author{Hongyu Zhang}
\authornote{Corresponding author.}
\email{hongyujohn@gmail.com}
\affiliation{%
  \institution{Chongqing University}
  \country{China}
}

% Short author list for page headers
\renewcommand{\shortauthors}{Pan et al.}

% =========================================================
% Abstract
% =========================================================

\begin{abstract}

Generating deployable websites from visual prototypes and textual requirements is increasingly feasible with multimodal models, but the generated project must stay consistent with the prototypes it is built from.
In this setting, the prototypes specify page layouts and visual elements, while the requirements describe interaction workflows and deployment constraints.
Existing UI-to-code methods and coding agents often handle layout reconstruction, interaction implementation, and project-file editing as separate stages.
Because these stages are separate, visual prototype elements, project files, interaction workflows, dependency relations, and render evidence can lose their correspondence across generation, verification, and update, which we call \textit{prototype-to-project state drift}.
In this paper, we propose \approach, a state-aware framework that mitigates this drift by maintaining these correspondences in a Project State Graph.
The graph records typed links among four types of state nodes and their dependency relations, and serves as the shared representation for prototype grounding, graph-guided project generation, and render-evidence-guided scoped update, which maps failed checks back to the affected project files and leaves files outside the update scope unchanged.
We evaluate \approach on Vision2Web with five foundation models across Static Webpage, Interactive Frontend, and Full-Stack Website tasks.
Averaged across the five models, \approach improves the benchmark's overall score over the strongest baseline under each model by 26.1 and 22.6 points on Interactive Frontend and Full-Stack Website tasks, respectively.
To measure drift directly, we track how often checks that pass in one update round fail in a later round, and find that \approach reduces verified-state drift compared with alternative state representations.
An annotation study on 50 tasks shows that the constructed graphs reach an edge-level F1 of 95.6\%.

\end{abstract}

% =========================================================
% CCS Concepts
% =========================================================




% =========================================================
% Generate Title
% =========================================================

\maketitle

% =========================================================
% Main Sections
% =========================================================

\section{Introduction}

Generating deployable websites from visual prototypes and textual requirements is an important step toward automated web development~\cite{si2025design2code,gui2025uicopilot,nguyen2015reverse,moran2018machine,wan2025divide,wu2025mllm}.
In this setting, visual prototypes specify page layouts and prototype elements, while textual requirements describe the interaction workflows and deployment requirements that the generated project must implement.
Multimodal large language models, UI-to-code methods, and coding agents have made this task increasingly feasible~\cite{si2025design2code,yun2024web2code,xiao2025interaction2code,he2026vision2web,yang2024swe,wang2025openhands,anthropic2026claudecode}, but many of them handle layout reconstruction, interaction implementation, and project-file editing as separate stages.
As a result, they may produce plausible pages, interactions, or file edits that do not form a consistent project state across generation, verification, and update.

\textbf{The central challenge is \textit{prototype-to-project state drift}.}
For example, a prototype element may be implemented through multiple project files, styles, and assets.
An interaction workflow may involve a visual trigger, an event handler, a route, and a target page.
A render or interaction failure may expose the symptom, but the cause may lie in an incorrect link between the prototype element, the implementing file, the workflow, or the dependency relation~\cite{moran2018automated,choudhary2013x}.
When such links are not maintained, drift surfaces in two observable ways.
First, correspondences break: the prototype element that should start a workflow is rendered and operable, but the workflow fails.
Second, verified states regress: a check that passes in one update round fails in a later round.
Section~\ref{sec:motivating} turns both symptoms into metrics, which we use to evaluate whether the Project State Graph reduces drift rather than only improving final scores.

Existing approaches often do not explicitly maintain persistent prototype-to-project correspondences throughout generation, verification, and update.
\ding{182} \textbf{Static UI-to-code methods} can reconstruct page layouts and components from visual prototypes, but they usually do not link these visual outputs to interactions, project files, and render evidence~\cite{jiang2025screencoder,gui2025uicopilot,chen2020object}.
\ding{183} \textbf{Interactive UI generation methods} go beyond static layout reconstruction by implementing user interactions and workflows, but they usually do not preserve the correspondence among these interactions, prototype elements, project files, and render evidence during iterative updates~\cite{xu2025webvia}.
\ding{184} \textbf{General coding agents} can edit project files and incorporate tool feedback, but they are not designed to align these edits with prototype elements or to map render evidence back to the affected project files and interaction workflows~\cite{yang2024swe,wang2025openhands}.
\ding{185} \textbf{Specification-to-repository frameworks}, such as ARC~\cite{kong2026arc}, trace requirement documents to code and verify them with synthesized tests, so a mismatch between the rendered pages and the visual prototypes is not traced back to the prototype element and project files it concerns; Section~\ref{sec:related} compares the two settings directly.
Because none of these methods keeps prototype-to-project correspondences as explicit project state, they remain exposed to prototype-to-project state drift.

\textbf{In this paper, we propose \approach, a state-aware framework that mitigates \textit{prototype-to-project state drift} by maintaining prototype-to-project correspondences in a Project State Graph.}
The Project State Graph records typed links among visual prototype elements, project files, interaction workflows, and render evidence, including their dependency relations, so that each generation or update step can query the affected visual, code, workflow, and evidence states together.
This shared state representation guides three operations.
First, prototype grounding aligns visual prototypes with textual requirements and initializes prototype-to-project correspondences.
Second, graph-guided project generation uses the graph to generate project files under visual, interaction, and dependency constraints.
Third, render-evidence-guided scoped update maps render evidence back to affected project states and revises only the project files in the affected scope.
This design builds on two established software engineering techniques: the links of the graph act as trace links in requirements traceability~\cite{cleland2014traceability}, and the update scope is a reachability-based impact set over dependency relations, as in change impact analysis~\cite{li2013survey} and dependency-graph-based repository editing~\cite{bairi2024codeplan}.
\approach adapts these techniques to deployable website generation in three respects:
(i) trace links are fixed by the implementation plan before project files are produced, rather than recovered afterward, and they connect visual prototype elements and requirement-derived workflows to code;
(ii) render evidence enters the graph as nodes, so impact analysis starts from an observed structural or visual failure on the rendered pages rather than from a code change;
(iii) the update is write-protected outside the impact set, so files not implicated by a failed check remain unchanged.

We evaluate \approach on the Vision2Web benchmark~\cite{he2026vision2web} with five foundation models across Static Webpage, Interactive Frontend, and Full-Stack Website tasks.
Averaged across the five models, \approach improves the benchmark's overall score over the strongest baseline under each model by 26.1 points on Interactive Frontend tasks and 22.6 points on Full-Stack Website tasks, where a project spans multiple pages, workflows, and files.
The improvement coincides with less drift: compared with the strongest alternative, \approach lowers the correspondence break rate from 44.2\% to 33.3\% and the verified-state drift rate from 3.0\% to 1.1\%.
An annotation study on 50 tasks shows that the constructed graphs achieve an edge-level F1 of 95.6\%.

In summary, this paper makes the following contributions:
\begin{itemize}

\item We identify \textit{prototype-to-project state drift} as a key challenge in deployable website generation and make it measurable through two observable symptoms, broken correspondences and regressions of verified states, quantified by the correspondence break rate and the verified-state drift rate.

\item We propose \approach, which maintains prototype-to-project correspondences in a Project State Graph.
The graph adapts trace links and reachability-based impact analysis to deployable website generation: it fixes cross-modal links before code is written, starts impact analysis from render evidence, and write-protects files outside the update scope.

\item We evaluate \approach on Vision2Web with five foundation models and three task levels, including direct drift measurement, an annotation study of graph correctness, and a cost analysis.
We additionally evaluate \approach on WebGen-Bench~\cite{lu2025webgenbench}, where no visual prototypes are available.
We release the implementation, prompts, generated projects, evaluation logs, and annotation data.

\end{itemize}

\section{Motivating Example}
\label{sec:motivating}

\begin{wrapfigure}{r}{0.48\linewidth}
    \centering
    \vspace{-8pt}
    \includegraphics[width=\linewidth]{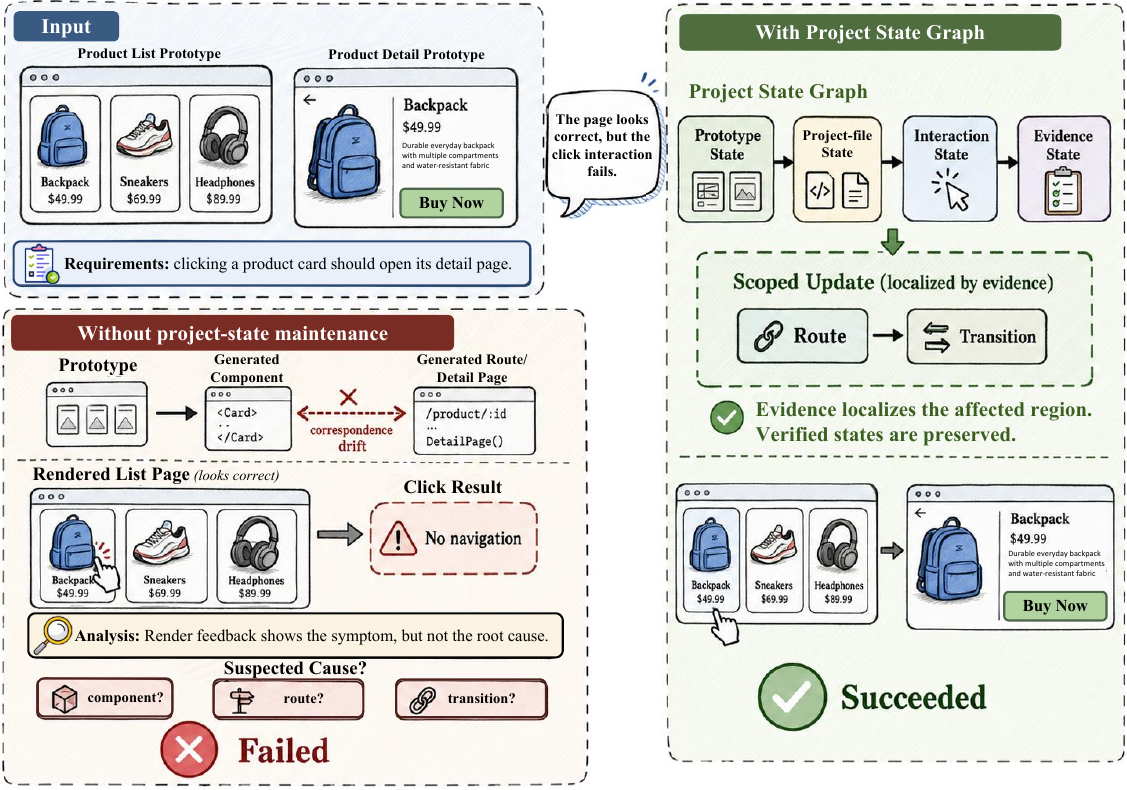}
    \caption{Motivating example of prototype-to-project state drift.}
    \label{fig:motivating}
    \vspace{-10pt}
\end{wrapfigure}

Figure~\ref{fig:motivating} shows a task with two prototypes, a product-list page and a product-detail page, and the requirement that clicking a product card opens the corresponding detail page.
A model may generate both pages so that each matches its prototype, yet leave the product card, its click handler, and the detail-page route unconnected: the list page looks correct, but the click triggers no navigation or opens the wrong page.

This failure is hard to repair after generation.
A screenshot of the list page matches the prototype, so the broken navigation does not appear in render evidence, and the link among the card, its handler, and the route has to be established when the project is generated.
For defects that render evidence does reveal, such as a missing product image, the evidence shows the symptom but not the files to change, since the card may be implemented across the page, its stylesheet, and a shared theme.
\approach links the card to its click workflow and the detail-page route before any code is written, and maps render evidence back to the files a visual defect concerns.

\noindent\textbf{Measuring drift.}
Drift cannot be observed directly, but its two symptoms can.
We compute both from the benchmark's official functional test cases and evaluator, run on the final and on every intermediate project, so they do not depend on the checks of \approach.
For a task $\tau$ with test cases $C_\tau$, let $\mathit{pass}(c)$ indicate that test case $c$ passes (failed and blocked cases count as not passed), and let $\mathit{reach}(c)$ indicate that, when the evaluator reaches $c$ in the official execution order, it locates the element of the first action of $c$ and completes that action.
The \textit{correspondence break rate} measures failures that occur after the trigger is already rendered and operable, i.e., failures downstream of the trigger, in its handler, route, target state, or backend, rather than in rendering:
\begin{equation}
\mathrm{CBR}=\frac{\sum_{\tau}|\{c \in C_\tau \mid \mathit{reach}(c) \wedge \neg\mathit{pass}(c)\}|}{\sum_{\tau}|\{c \in C_\tau \mid \mathit{reach}(c)\}|},
\end{equation}
reported with the trigger-reachable rate $\mathrm{TRR}=\sum_{\tau}|\{c \mid \mathit{reach}(c)\}| / \sum_{\tau}|C_\tau|$, since rendering fewer triggers would lower CBR.
The \textit{verified-state drift rate} measures regressions: with $\mathit{pass}_t(c)$ evaluated on the project after the $t$-th of $T_\tau$ update rounds ($t=0$ after generation),
\begin{equation}
\mathrm{VSDR}=\frac{\sum_{\tau}\sum_{t<T_\tau}|\{c \mid \mathit{pass}_t(c) \wedge \neg\mathit{pass}_{t+1}(c)\}|}{\sum_{\tau}\sum_{t<T_\tau}|\{c \mid \mathit{pass}_t(c)\}|},
\end{equation}
reported with the repair rate RR, defined symmetrically as the fraction of failing test cases that pass after the next update, since a method that never edits the project has a VSDR of zero.
All four metrics are pooled over test cases, so a task whose update loop stops early contributes only its completed rounds.

\section{\approach}
\label{sec:approach}

\subsection{Overview}
\label{sec:approach_overview}

\begin{figure*}[t]
\centering
\includegraphics[width=0.75\textwidth]{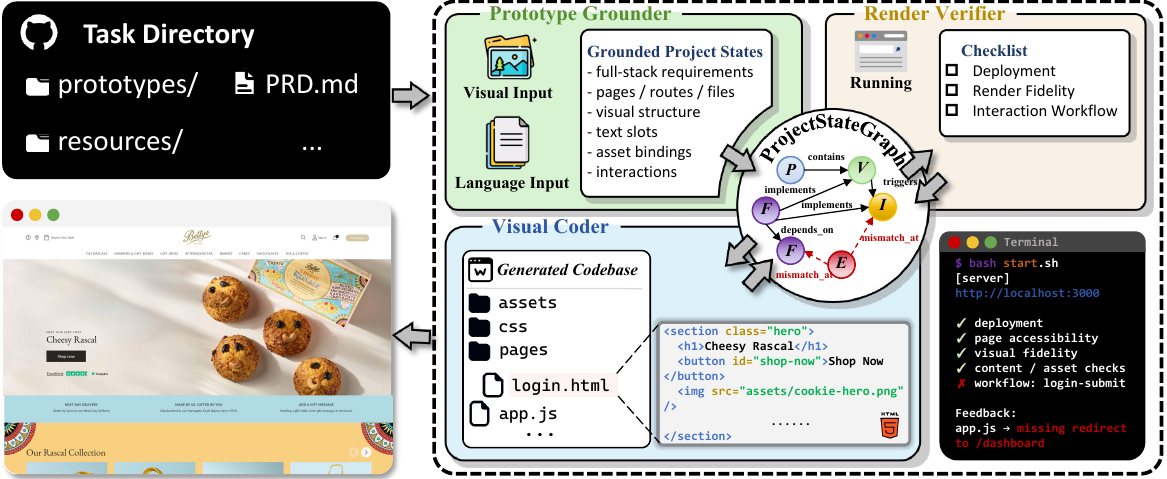}
\caption{
Overview of \approach.
}
\label{fig:overview}
\vspace{-10px}
\end{figure*}

Given visual prototypes $\mathcal{V}$ and textual requirements $\mathcal{R}$, \approach generates a deployable website project $\mathcal{P}$.
The generated project should preserve the page layout and visual elements specified by the prototypes, implement the interaction workflows described by the requirements, and be executable in the target deployment environment.
The key idea of \approach is to mitigate prototype-to-project state drift by maintaining prototype-to-project correspondences as an explicit project state throughout generation, verification, and update.
To this end, \approach builds a Project State Graph $\mathcal{G}$ that links prototype elements, project files, interaction workflows, and render evidence, including their dependency relations.
Figure~\ref{fig:overview} illustrates the workflow of \approach, and Algorithm~\ref{alg:visstatecoder} summarizes the corresponding procedure.

\begin{algorithm}[t]
\caption{State-aware deployable website generation in \approach}
\label{alg:visstatecoder}
\footnotesize
\begin{algorithmic}[1]
\Require Visual prototypes $\mathcal{V}$, textual requirements $\mathcal{R}$, maximum update rounds $K$
\Ensure Generated deployable website project $\mathcal{P}$

\State $(\mathcal{G}, \mathcal{T}) \gets \textsc{PrototypeGrounding}(\mathcal{V}, \mathcal{R})$

\For{each prototype-grounded task $\tau \in \mathcal{T}$}
    \State $\mathcal{S}_{\tau} \gets \textsc{ExtractTaskSubgraph}(\mathcal{G}, \tau)$
    \State $\mathcal{C}_{\tau} \gets \textsc{BuildGenerationContext}(\mathcal{S}_{\tau}, \tau)$
    \State $\Delta\mathcal{F}_{\tau} \gets \textsc{GenerateProjectFiles}(\mathcal{C}_{\tau})$
    \State $\mathcal{G} \gets \textsc{UpdateGraphWithProjectFiles}(\mathcal{G}, \Delta\mathcal{F}_{\tau})$
\EndFor

\State $\mathcal{P} \gets \textsc{AssembleProject}(\mathcal{G})$

\For{$i \gets 1$ to $K$}
    \State $\mathcal{O}_{i} \gets \textsc{CollectRenderEvidence}(\mathcal{P}, \mathcal{V}, \mathcal{R})$
    \State $\mathcal{G} \gets \textsc{UpdateGraphWithRenderEvidence}(\mathcal{G}, \mathcal{O}_{i})$
    \If{$\textsc{AllChecksPass}(\mathcal{O}_{i})$}
        \State \textbf{break}
    \EndIf
    \State $\Omega_0 \gets \textsc{LocalizeMismatches}(\mathcal{G}, \mathcal{O}_{i})$
    \State $\Omega \gets \textsc{BuildUpdateScope}(\mathcal{G}, \Omega_0)$
    \State $\mathcal{P} \gets \textsc{ApplyScopedUpdate}(\mathcal{P}, \Omega, \mathcal{O}_{i})$
    \State $\mathcal{G} \gets \textsc{UpdateGraphWithScopedUpdate}(\mathcal{G}, \mathcal{P}, \Omega, \mathcal{O}_{i})$
\EndFor

\State \Return $\mathcal{P}$
\end{algorithmic}
\end{algorithm}

\subsection{Project State Graph}
\label{sec:psg}

\noindent\textbf{Graph schema.}
We define the Project State Graph as $\mathcal{G}=(\mathcal{N},\mathcal{E},\mathcal{A})$, where $\mathcal{N}$ denotes typed state nodes, $\mathcal{E}$ denotes typed relation edges, and $\mathcal{A}$ stores node and edge attributes.
Table~\ref{tab:psg_schema} lists its four types of state nodes and seven types of edges.

\begin{table}[t]
\centering
\caption{Schema of the Project State Graph.}
\label{tab:psg_schema}
\footnotesize
\setlength{\tabcolsep}{3pt}
\renewcommand{\arraystretch}{1.10}
\begin{tabular}{p{0.21\linewidth}p{0.36\linewidth}p{0.34\linewidth}}
\toprule
\textbf{Element} & \textbf{Meaning} & \textbf{Key attributes / relations} \\
\midrule

Prototype state &
Visual page, UI region, or visual prototype element. &
Page or region id, role, reading order, bounding box, style hints, text and asset slots. \\

Project-file state &
Generated file, component, route, asset, or configuration in the project. &
File path, purpose, file type, implementation status, rendered region anchors. \\

Interaction state &
User workflow specified by textual requirements. &
Source page, trigger and target text, event type, target route or state, required inputs, expected texts. \\

Render-evidence state &
Structural, pixel, or visual-judge observation collected during verification. &
Pass/fail status, issues with symptom class and localized identifiers, screenshots. \\

\midrule
\texttt{contains} &
Hierarchy among pages, regions, and slots. &
Page $\rightarrow$ region, interaction; region $\rightarrow$ slot; page file $\rightarrow$ rendered region. \\

\texttt{grounds\_to} / \texttt{triggers} &
Link an interaction and the prototype element that initiates it, in both directions. &
Interaction $\leftrightarrow$ prototype state. \\

\texttt{implements} &
Links a planned page or route to the file that implements it. &
Page or route $\rightarrow$ project file. \\

\texttt{depends\_on} &
Records a planned dependency. &
Route $\rightarrow$ page; asset slot $\rightarrow$ asset; interaction $\rightarrow$ target route. \\

\texttt{verified\_by} &
Links states to the evidence of a round in which all checks pass. &
Page, file, or interaction $\rightarrow$ render evidence. \\

\texttt{mismatch\_at} &
Links failed evidence to the states it concerns. &
Render evidence or issue $\rightarrow$ localized state. \\

\bottomrule
\end{tabular}
\vspace{-10px}
\end{table}

\noindent\textbf{Graph update.}
The Project State Graph evolves across generation, verification, and update.
Prototype grounding creates the initial graph and task set.
The \texttt{implements} and \texttt{depends\_on} edges are derived from the implementation plan produced at the end of prototype grounding, which assigns every page and route to a project file and binds asset slots to resource files.
Because these assignments are fixed before code is written, each generation task produces exactly its assigned file, and the links do not have to be recovered from the generated code.
After generation and after every update, \approach scans the workspace, refreshes the project-file states, and parses the \texttt{data-region-id} anchors of each HTML file into \texttt{contains} edges from the file to its rendered regions.
\approach does not parse imports between source files, so a file shared by several pages, such as a common script, is not connected by a dependency edge; the file-level update scope in Section~\ref{sec:scoped_update} covers such files through the page that reports the failure.
After verification, passing render evidence is connected to verified graph states through \texttt{verified\_by} edges, while failing render evidence is connected to affected graph states through \texttt{mismatch\_at} edges (Section~\ref{sec:scoped_update}).

\subsection{Prototype Grounding}
\label{sec:prototype_grounding}

Prototype grounding converts the visual prototypes and textual requirements into the initial graph through a sequence of structured MLLM calls whose JSON outputs are validated against typed schemas and re-requested when validation fails.

\noindent\textbf{Visual grounding.}
For each prototype image, the model first extracts a page skeleton: the page with its route and file name, and its UI regions with a role, a reading order, a normalized bounding box, and a summary.
Subsequent calls are conditioned on the skeleton and may only use its page and region identifiers.
They extract the required text slots and asset slots of each region, a structure brief describing hierarchy and repeated elements, region-level style hints for layout, spacing, alignment, typography, and color, and global style tokens.
The bounding boxes, reading order, and style hints of a region serve as its layout constraints; they are soft constraints passed to generation and verification rather than a formal layout specification.
A separate call binds asset slots to local resource files, using the prototype and contact sheets of the available resources.

\noindent\textbf{Workflow extraction.}
Interaction workflows are extracted from the textual requirements, i.e., the task description for Interactive Frontend tasks and the requirement document for Full-Stack Website tasks, together with the page skeleton.
Each workflow becomes an interaction state with a source page, a trigger description and target text, an event type, a target route or target state, required inputs, and texts expected after the interaction.
A further call decides whether the requirements need a backend by combining keyword evidence, such as login, cart, or persistence, with the model's judgment, and if so plans data models and API routes.

\noindent\textbf{Element matching.}
An interaction state is linked to prototype elements by matching its trigger text against the text, name, role, and summary attributes of the text-slot, region, and page states of its source page.
Each matched state receives a \texttt{grounds\_to} edge from the interaction state and a \texttt{triggers} edge to it.
If no element matches, the interaction is grounded to its source page, which keeps it reachable in later updates at page granularity.

\noindent\textbf{Graph validation.}
Grounding enforces referential consistency: generated identifiers must exist in the skeleton, interactions with an unknown source page are discarded, and localization returns only existing states.
It does not check whether the extracted states are semantically consistent with the inputs; Section~\ref{sec:rq4} measures the accuracy of the resulting graph and how its errors propagate.

Prototype grounding also derives a set of prototype-grounded tasks $\mathcal{T}$ from a final implementation plan, which assigns each page to an HTML file and route and lists the stylesheets, scripts, and, for full-stack projects, the server and data files of the project.
Each task $\tau \in \mathcal{T}$ produces one planned file, and its relevant graph states are those of the page, regions, text and asset slots, and interactions that the file implements.
Formally, prototype grounding produces:
\begin{equation}
(\mathcal{G}_0, \mathcal{T}) = \mathrm{Ground}(\mathcal{V}, \mathcal{R}).
\end{equation}

\subsection{Graph-Guided Project Generation}
\label{sec:graph_guided_generation}

For each prototype-grounded task $\tau \in \mathcal{T}$, \approach extracts a task-relevant subgraph from the Project State Graph and serializes it into a structured generation context.
For a page file, the subgraph is scoped to the page: it contains the page state with its structure brief, the text slots, asset slots, and asset bindings of its regions, its verification targets, and the interaction states whose source is the page, and the context is paired with the page's prototype image.
Files shared across pages are generated from the project-wide view after the pages they serve: the shared theme stylesheet from all generated pages and the global style tokens, the interaction script from all interaction states and the generated pages, whose region anchors it must attach to, and, when a backend is required, the server from the planned data models and API routes.
Per-page stylesheets combine the page-scoped subgraph with the generated page and the shared theme.
Start scripts and package manifests are generated deterministically.
The context specifies which prototype elements to match, which interaction workflows to implement, and which planned regions and texts the file must contain, and the generation prompt forbids sections, texts, assets, and routes that are not in the graph.
Based on this context, \approach generates the corresponding project file, and each written file is recorded as an implemented project-file state.

\subsection{Render-Evidence-Guided Scoped Update}
\label{sec:scoped_update}

\noindent\textbf{Render evidence.}
After the project $\mathcal{P}_i$ is assembled from the project files represented in the Project State Graph, \approach collects three kinds of render evidence in each update round without using the benchmark's test cases or evaluator.
(1) \textit{Structural checks} verify that every planned file exists, that the start script is executable, and that each page links its stylesheets and scripts, contains its planned region anchors, required texts, and bound assets, and renders no placeholder blocks, unplanned sections or routes, or remote assets.
(2) \textit{Pixel comparison} renders each page with Playwright at its prototype viewport and reports a visual mismatch when the root-mean-square difference to the prototype exceeds 0.28.
(3) A \textit{visual judge}, which follows the component-level protocol of the VS metric (Section~\ref{sec:metrics}) and is implemented with the same foundation model, segments the prototype and the screenshot into components, scores each component on a five-level scale, and reports components scoring below 0.75 with a mismatch category.
Each failed check becomes an issue that carries the identifiers it concerns: the page file and its stylesheets and, when available, the region, text slot, asset slot, or component.
Verification does not execute interaction workflows, since a broken click or navigation is not visible in a screenshot; \approach addresses interaction correctness when the project is generated, through the \texttt{grounds\_to}, \texttt{triggers}, and \texttt{depends\_on} links that tie each trigger element to its workflow and target route.

\noindent\textbf{Localization.}
\approach inserts the collected evidence into the Project State Graph as a render-evidence state.
Localization maps the identifiers of each issue to graph states by lookup rather than by inference: a missing file maps to its project-file state, a missing text, asset, or region anchor to the corresponding slot or region state, and a pixel or visual-judge mismatch to the page and its files.
\approach adds a \texttt{mismatch\_at} edge from the render-evidence state and from the issue to every matched state, and groups issues that concern the same file so that one update addresses them together.

Given a failed render-evidence state $o_f$, \approach first identifies the directly affected prototype, project-file, or interaction states:
\begin{equation}
\Omega_0
=
\{ n \mid (o_f, \texttt{mismatch\_at}, n) \in \mathcal{E} \}.
\end{equation}
To include the files and relations required for a valid update, \approach expands $\Omega_0$ to its two-hop neighborhood over \texttt{implements}, \texttt{triggers}, and \texttt{depends\_on} edges, traversed in either direction:
\begin{equation}
\Omega = \mathcal{N}^{\le 2}_{\mathcal{G}}\big(\Omega_0,\{\texttt{implements},\texttt{triggers},\texttt{depends\_on}\}\big).
\end{equation}
Bounding the expansion to two hops reaches the files implementing an affected page or route and the routes and pages an affected interaction depends on, without pulling in the rest of the project.
If localization matches no state, $\Omega$ falls back to at most five project-file states, which bounds the patch size when the evidence carries no identifier that resolves to a state.

\noindent\textbf{Scoped update.}
The update is enforced at the file level.
A patch may write only the files of project-file states in $\Omega$ and the page, page stylesheet, and shared theme files named by the grouped issues; the patch validator rejects any patch that writes another file.
Files outside this set, including those whose checks passed in earlier rounds, therefore remain unchanged.
Within the allowed files, preservation relies on the model: the update prompt contains the issue, the affected subgraph, and the current contents of the allowed files, and asks for a minimal patch that preserves non-target regions.
Updates are applied in three stages: deterministic fixes for missing links and files, model-generated patches for missing text, assets, regions, and placeholders, and CSS-only patches for visual mismatches, which may only append rules to one stylesheet.
After the update, \approach re-verifies the project, marks issues that no longer appear as resolved, and, when all checks pass, links every page, file, interaction, and verification-target state to the passing evidence through \texttt{verified\_by} edges and stops.
\approach does not roll back an update that breaks a previously passing check; the next round reports the regression as a new issue, and the verified-state drift rate in RQ2 measures how often this happens.

\section{Experimental Setup}

We organize the evaluation around the following research questions (RQs):

\begin{itemize}

    \item \textbf{RQ1. How does \approach compare with representative deployable website generation baselines?}
    We evaluate the end-to-end effectiveness of \approach on Vision2Web across Static Webpage, Interactive Frontend, and Full-Stack Website tasks with five foundation models.

    \item \textbf{RQ2. To what extent does the Project State Graph mitigate prototype-to-project state drift?}
    We compare \approach with three variants that keep the rest of the framework unchanged: \textbf{w/o PSG}, which removes the shared project state; \textbf{Text-State Memory (TSM)}, which records the project state in natural language; and \textbf{JSON-State}, which stores the same information as the graph in a structured JSON record without typed graph operations.
    Besides the benchmark scores, we measure the two drift symptoms defined in Section~\ref{sec:motivating}.

    \item \textbf{RQ3. How do prototype grounding and render-evidence-guided scoped update contribute to \approach?}
    We ablate the two operations that create and use the graph: prototype grounding before generation and render-evidence-guided scoped update after generation.

    \item \textbf{RQ4. How accurate is the constructed Project State Graph, and how do construction errors propagate?}
    We annotate the nodes and edges of the graphs built for a sample of tasks and relate construction errors to the final functional results.

\end{itemize}

\subsection{Benchmark}

We evaluate on Vision2Web~\cite{he2026vision2web}, which is built from real-world websites in 16 subcategories across four domains.
It contains 100 Static Webpage tasks, each with desktop, tablet, and mobile prototypes of one page; 66 Interactive Frontend tasks with, on average, 5.9 prototypes, an interaction description of 1.0K tokens, and 7.5 test cases; and 27 Full-Stack Website tasks with, on average, 8.5 prototypes, a requirement document of 4.3K tokens that also covers state management and backend behavior, and 28.2 test cases~\cite{he2026vision2web}.
In total, the benchmark has 918 prototypes and 1{,}255 functional test cases.

\subsection{Metrics}
\label{sec:metrics}

All methods are scored with the official Vision2Web evaluator, which deploys each project with its start script in the benchmark's Docker sandbox; a project that fails to deploy within ten minutes receives zero on all its prototypes and test cases.
\ding{182} \textbf{Visual Score (VS).}
A VLM judge splits each prototype and the screenshot of the corresponding page into functional components, such as the header and the main content, and scores each component on a five-level scale from 0 to 1.
A prototype's score is the mean over its components, and VS averages all prototypes of a level (0--100); for Static Webpage tasks, we also report VS per device.
\ding{183} \textbf{Functional Score (FS).}
FS is the percentage of expert-written test cases that pass, with failed and blocked cases counted as not passed.
A GUI agent runs the test cases of each workflow in one browser session without access to the source code; each test case performs user actions, such as clicks, text input, selections, back navigation, and refresh, and validates the rendered content, the URL, client-side storage, and HTTP API responses.
Because each test case starts from the state left by the previous one, FS covers form handling, state updates, persistence, and backend behavior in addition to navigation, i.e., workflow-level functional correctness; WebGen-Bench~\cite{lu2025webgenbench} uses the same agent-executed paradigm.
\ding{184} \textbf{Average Score (Avg)} is $\mathrm{VS}$ for Static Webpage tasks, equal to the mean of the three device scores, and $(\mathrm{VS}+\mathrm{FS})/2$ otherwise, computed from the pooled values of the level.
Following the official aggregation, VS and FS are pooled over all prototypes and test cases of a level rather than averaged per task.
We also report the \textbf{Deployment Success Rate (DSR)}, which Vision2Web lists as a reference indicator, and, for RQ2, the drift metrics CBR/TRR and VSDR/RR of Section~\ref{sec:motivating}, computed from the same test-case results.

\subsection{Baselines}

We compare \approach with three groups of baselines, each run with the same five foundation models.
For all of them, we follow the released code, documentation, or prompts when available; none uses the Project State Graph or scoped update.

\textbf{Static UI-to-code methods.}
UICopilot~\cite{gui2025uicopilot} generates code hierarchically from a coarse HTML structure to fine-grained code, UIOrchestra~\cite{yue2025uiorchestra} iterates layout description, code generation, and difference analysis, and ScreenCoder~\cite{jiang2025screencoder} separates grounding, planning, and generation.
They take the prototypes as input and are evaluated standalone on Static Webpage tasks.
Because they produce no interactions, routes, or backend, we evaluate them at the other two levels as two-stage pipelines, denoted \textbf{+CC}: the static method generates the initial project, and a Claude Code completion agent, configured as in Section~\ref{sec:impl} and given the textual requirements and the generated files, adds the missing functionality.
The three pipelines share the completion agent, so their differences reflect the initial static project rather than the interaction capability of the static methods; we also report the static stage alone (Section~\ref{sec:rq1}).

\textbf{Interactive UI-to-code method.}
WebVIA~\cite{xu2025webvia} explores, generates, and validates interactive interfaces; it receives the prototypes and the textual requirements.

\textbf{Coding agent.}
Claude Code~\cite{anthropic2026claudecode} receives the same prototypes, resources, and requirements as \approach and builds, runs, and edits the project with its own tools.

\subsection{Implementation Details}
\label{sec:impl}

We evaluate \approach with five foundation models: \textbf{Doubao-Seed-1.8, Gemini-3-Flash-Preview, Gemini-3.1-Pro-Preview, GPT-5.1, and Qwen3.5-Plus}.
The same model serves all model calls of a run, including the visual judge used inside \approach.
\approach calls each model through an OpenAI-compatible chat-completions endpoint with temperature 0.2, JSON output mode, and at most 16{,}384 output tokens per call.
Baselines use the decoding settings of their released implementations.

For each task, \approach takes the visual prototypes, resources, and textual requirements as input.
It runs at most $K=4$ update rounds; each round verifies the project, localizes the failed checks, applies a scoped update, and re-verifies the project, and the loop stops early when all checks pass or no update applies.
All Claude Code results, including those of the completion agent, are produced in our environment with the same five foundation models used for \approach.
Following the configuration of Vision2Web~\cite{he2026vision2web}, we run Claude Code v2.1.284 in non-interactive mode inside the benchmark's Docker sandbox, with one session per task, a per-task time limit of 7{,}200 seconds, and up to two retries when a session exits without producing a deployment script. Claude Code receives the same visual prototypes, resources, and textual requirements as \approach, and neither Claude Code nor \approach accesses the benchmark's test workflows or official evaluator during generation.

\section{Experimental Results}

\subsection{RQ1: Performance Comparison}
\label{sec:rq1}

To answer RQ1, we compare \approach with representative deployable website generation baselines on Vision2Web.
The comparison covers five foundation models and three task levels: Static Webpage, Interactive Frontend, and Full-Stack Website.
Table~\ref{tab:rq1_main_results} reports the detailed results.

\begin{table*}[!t]
\centering
% [Original caption: Performance comparison on Vision2Web. For UICopilot, UIOrchestra, and ScreenCoder, Static Webpage results are produced by the method alone; on Interactive Frontend and Full-Stack Website tasks, (+CC) denotes that their static project is completed by the Claude Code completion agent with the same foundation model (Section~\ref{sec:impl}).]
\caption{Performance comparison on Vision2Web. For UICopilot, UIOrchestra, and ScreenCoder, Static Webpage results are produced by the method alone; on the other two levels, (+CC) denotes that their static project is completed by the Claude Code completion agent with the same foundation model (Section~\ref{sec:impl}). Avg is the mean of the three device scores on Static Webpage tasks and $(\mathrm{VS}+\mathrm{FS})/2$ elsewhere.}
\label{tab:rq1_main_results}
\setlength{\tabcolsep}{4pt}
\renewcommand{\arraystretch}{1.08}

\definecolor{DeepBlue}{RGB}{77,88,234}
\definecolor{LightBlue}{RGB}{235,241,250}

\resizebox{0.82\textwidth}{!}{
\begin{tabular}{l|ccccccccccccc}
\toprule
\multirow{2}{*}{\textbf{Method}} 
& \multicolumn{5}{c}{\textbf{Static Webpage}} 
& \multicolumn{4}{c}{\textbf{Interactive Frontend}} 
& \multicolumn{4}{c}{\textbf{Full-Stack Website}} \\
\cmidrule(lr){2-6} \cmidrule(lr){7-10} \cmidrule(lr){11-14}
& \textbf{Desktop} 
& \textbf{Tablet} 
& \textbf{Mobile} 
& \textbf{Avg} 
& \textbf{DSR}
& \textbf{VS} 
& \textbf{FS} 
& \textbf{Avg} 
& \textbf{DSR}
& \textbf{VS} 
& \textbf{FS} 
& \textbf{Avg} 
& \textbf{DSR} \\
\midrule

\multicolumn{14}{c}{\textbf{\textit{Doubao-Seed-1.8}}} \\
\midrule
UICopilot\,(+CC)     & 58.6  & 59.7  & 57.5  & 58.6  & 92\%  & 20.3  & 27.5  & 23.9  & 92.4\%  & 16.8  & 10.4  & 13.6  & 77.8\%  \\
UIOrchestra\,(+CC)   & 58.2  & 56.3  & 56.0  & 56.8  & 97\%  & 23.2  & 31.6  & 27.4  & 93.9\%  & 20.8  & 13.7  & 17.3  & 81.5\%  \\
ScreenCoder\,(+CC)   & 32.8  & 30.4  & 29.8  & 31.0  & 97\%  & 15.3  & 20.5  & 17.9  & 90.9\%  & 9.3  & 17.8  & 13.6  & 74.1\%  \\
WebVIA        & 53.2  & 55.6  & 54.9  & 54.6  & 99\%  & 17.9  & 36.4  & 27.1  & 92.4\%  & 17.6  & 10.6  & 14.1  & 77.8\%  \\
% [Avg cells corrected: Static Avg 17.3$\rightarrow$17.0, Interactive Avg 9.6\%$\rightarrow$10.3 (\% also removed). Original row kept for reference:
% Claude Code   & 19.3 & 16.9 & 14.8 & 17.3 & 39\% & 6.7 & 13.8 & 9.6\% & 34.4\% & 0 & 0 & 0 & 0\% \\ ]
Claude Code   & 19.3 & 16.9 & 14.8 & 17.0 & 39\% & 6.7 & 13.8 & 10.3 & 34.4\% & 0 & 0 & 0 & 0\% \\
\rowcolor{LightBlue}
\textcolor{DeepBlue}{\textbf{\approach}} 
& 63.7  & 57.5  & 58.1  & 59.8  & 100\%  & 50.6  & 55.7  & 53.2  & 98.5\%  & 35.6  & 40.3  & 38.0  & 100\%  \\
\midrule

\multicolumn{14}{c}{\textbf{\textit{Gemini-3-Flash-Preview}}} \\
\midrule
UICopilot\,(+CC)     & 56.2  & 63.8  & 65.5  & 61.8  & 94\%  & 28.7  & 34.9  & 31.8  & 89.4\%  & 20.7  & 17.3  & 19.0  & 85.2\%  \\
UIOrchestra\,(+CC)   & 58.9  & 57.4  & 59.2  & 58.5  & 98\%  & 34.7  & 35.9  & 35.3  & 97\%  & 26.4  & 17.8  & 22.1  & 85.2\%  \\
ScreenCoder\,(+CC)   & 12.1  & 13.4  & 11.9  & 12.5  & 96\%  & 6.5  & 18.4  & 12.5  & 87.9\%   & 9.8  & 17.0  & 13.4  & 70.4\%  \\
WebVIA        & 60.2  & 62.5  & 66.2  & 63.0  & 98\%  & 30.4  & 39.5  & 35.0  & 95.5\%  & 22.4  & 18.9  & 20.6  & 81.5\%  \\
Claude Code   & 48.5 & 41.7 & 39.7 & 43.3 & 92\% & 13.1 & 27.9 & 20.5 & 78.8\% & 2.3 & 4.6 & 3.5 & 63\% \\
\rowcolor{LightBlue}
\textcolor{DeepBlue}{\textbf{\approach}} 
& 65.0  & 66.4  & 68.9  & 66.8  & 100\%  & 53.4  & 72.2  & 62.8   & 100\%  & 51.0  & 40.8  & 45.9  & 96.3\%  \\
\midrule

\multicolumn{14}{c}{\textbf{\textit{Gemini-3.1-Pro-Preview}}} \\
\midrule
UICopilot\,(+CC)     & 60.2  & 59.3  & 58.4  & 59.3  & 94\%  & 29.9   & 41.6  & 35.8  & 87.8\%  & 24.0  & 20.8  & 22.4  & 85.2\%  \\
UIOrchestra\,(+CC)   & 60.2  & 61.1  & 61.6  & 61.0  & 99\%  & 34.0  & 40.5  & 37.3  & 97\%  & 27.6  & 19.4  & 23.5  & 88.9\%  \\
ScreenCoder\,(+CC)   & 8.3   & 8.0   & 11.4  & 9.2   & 100\%  & 7.9  & 21.7  & 14.8  & 89.4\%  & 6.4  & 16.7  & 11.6  & 66.7\%  \\
WebVIA        & 60.8  & 59.3  & 62.4  & 60.8  & 98\%  & 25.5  & 38.5  & 32.0  & 95.5\%  & 20.6  & 19.9  & 20.2  & 81.5\%  \\
% [Static Avg corrected: 53.3$\rightarrow$56.3. Original: Claude Code   & 65.7 & 53.2 & 50.1 & 53.3 & 94\% & 15.3 & 16.1 & 14.8 & 68.4\% & 7.3 & 14.7 & 8.1 & 67.2\% \\ ]
Claude Code   & 65.7 & 53.2 & 50.1 & 56.3 & 94\% & 15.3 & 16.1 & 15.7 & 68.4\% & 7.3 & 14.7 & 11.0 & 67.2\% \\
\rowcolor{LightBlue}
\textcolor{DeepBlue}{\textbf{\approach}} 
& 62.8  & 60.2  & 60.3  & 61.1  & 100\%  & 56.9  & 57.5  & 57.2  & 100\%  & 50.3  & 36.5  & 43.4  & 100\%  \\
\midrule

\multicolumn{14}{c}{\textbf{\textit{GPT-5.1}}} \\
\midrule
UICopilot\,(+CC)     & 63.6  & 59.4  & 53.8  & 58.9  & 96\%  & 35.3  & 50.3  & 42.8   & 89.4\%  & 29.7  & 26.5  & 28.1  & 88.9\%  \\
UIOrchestra\,(+CC)   & 57.6  & 60.2  & 61.3  & 59.7  & 97\%  & 42.7  & 51.7  & 47.2  & 98.5\%  & 33.5  & 25.9  & 29.7  & 88.9\%  \\
ScreenCoder\,(+CC)   & 27.2  & 6.7   & 23.0  & 19.0  & 98\%  & 12.8  & 36.7  & 24.8  & 95.5\%  & 9.5  & 24.1  & 16.8  & 81.5\%  \\
WebVIA        & 61.7  & 61.2  & 60.9  & 61.3  & 99\%  & 30.2  & 45.5  & 37.9  & 97\%  & 24.8  & 25.2  & 25.0  & 85.2\%  \\
% [Corrected: Static Avg 44.7$\rightarrow$44.5, Interactive Avg 43.8$\rightarrow$42.6, Full-Stack Avg 14.9$\rightarrow$15.7. Original: Claude Code   & 47.5 & 45.3 & 40.6 & 44.7 & 98\% & 22.4 & 62.7 & 43.8 & 97.4\% & 8.8 & 22.5 & 14.9 & 84.7\% \\ ]
Claude Code   & 47.5 & 45.3 & 40.6 & 44.5 & 98\% & 22.4 & 62.7 & 42.6 & 97.4\% & 8.8 & 22.5 & 15.7 & 84.7\% \\
\rowcolor{LightBlue}
\textcolor{DeepBlue}{\textbf{\approach}} 
& 61.2  & 63.5  & 60.4  & 61.7  & 100\%  & 57.6  & 80.9  & 69.3  & 100\%  & 50.1  & 50.8  & 50.4  & 100\%  \\
\midrule

\multicolumn{14}{c}{\textbf{\textit{Qwen3.5-Plus}}} \\
\midrule
UICopilot\,(+CC)     & 59.3  & 56.2  & 55.2  & 56.9  & 90\%  & 23.1  & 28.9  & 26.0  & 92.4\%  & 16.2  & 10.6  & 13.4  & 77.8\%  \\
UIOrchestra\,(+CC)   & 57.8  & 57.6  & 57.7  & 57.7  & 96\%  & 17.9  & 25.0  & 21.5  & 90.9\%  & 15.2  & 7.0  & 11.1  & 77.8\%  \\
ScreenCoder\,(+CC)   & 25.3  & 24.1  & 24.7  & 24.7  & 99\%  & 6.0  & 21.7  & 13.9  & 87.9\%  & 11.9  & 14.1  & 13.0  & 70.4\%  \\
WebVIA        & 60.1  & 57.9  & 58.6  & 58.9  & 95\%  & 7.9  & 26.5  & 17.2  & 87.9\%  & 10.5  & 7.3  & 8.9  & 70.4\%  \\
% [Corrected: Static Avg 34.3$\rightarrow$34.0, Interactive Avg 11.9$\rightarrow$11.7, Full-Stack Avg 7.7$\rightarrow$3.4. Original: Claude Code   & 33.2 & 37.1 & 31.6 & 34.3 & 87\% & 7.6 & 15.7 & 11.9 & 62.8\% & 1.6 & 5.1 & 7.7 & 62.8\% \\ ]
Claude Code   & 33.2 & 37.1 & 31.6 & 34.0 & 87\% & 7.6 & 15.7 & 11.7 & 62.8\% & 1.6 & 5.1 & 3.4 & 62.8\% \\
\rowcolor{LightBlue}
\textcolor{DeepBlue}{\textbf{\approach}} 
& 63.5  & 64.5  & 57.7  & 61.9  & 100\%  & 54.8  & 67.9  & 61.4  & 97\%  & 39.8  & 42.5  & 41.1  & 100\%  \\

\bottomrule
\end{tabular}
}
\vspace{-10px}
\end{table*}

Table~\ref{tab:rq1_main_results} shows that \approach achieves the best overall performance across the five foundation models and three task levels.
On Static Webpage tasks, \approach obtains the highest Avg score under all five models and the highest or tied highest DSR in all settings.
On Interactive Frontend tasks, its Avg ranges from 53.2 to 69.3, compared with 26.0 to 47.2 for the strongest baseline under each model.
On Full-Stack Website tasks, its Avg ranges from 38.0 to 50.4, compared with 13.4 to 29.7 for the strongest baseline.
Averaged across the five models, \approach improves Avg from 34.6 to 60.8 on Interactive Frontend tasks and from 21.2 to 43.8 on Full-Stack Website tasks, corresponding to gains of 26.1 and 22.6 points, respectively.

The gains become larger as tasks require more than static visual reconstruction.
Interactive Frontend and Full-Stack Website tasks additionally require interaction workflows, routing logic, and dependency relations to stay consistent, and the deployed project to be valid.
For the +CC pipelines, the static stages alone perform substantially worse on these tasks, so much of their final performance comes from the Claude Code completion stage rather than from the static method.
WebVIA and Claude Code can generate or modify interactive projects, but neither explicitly maintains correspondences among prototype elements, project files, workflows, dependencies, and render evidence.
In contrast, \approach maintains these relations through the Project State Graph, which helps preserve consistency as the generated website becomes more complex.

\begin{tcolorbox}[
  enhanced,
  colback=gray!15,
  colframe=gray!15,
  boxrule=0pt,
  arc=0pt,
  left=6pt,
  right=6pt,
  top=4pt,
  bottom=4pt,
  borderline west={2pt}{0pt}{gray!70}
]
\textbf{Answer to RQ1:}
\approach achieves the best overall performance under all five foundation models, with the largest gains on the two interactive task levels.
\end{tcolorbox}

\subsection{RQ2: Effect of the Project State Graph on Drift}
\label{sec:rq2}

To answer RQ2, we compare \approach with three variants that differ only in how project state is represented and used.
All variants use the same prototype grounding, generation prompts, verification checks, foundation models, decoding settings, and update budget.
\begin{itemize}
    \item \textbf{w/o PSG} removes the Project State Graph. Generation receives the grounding outputs directly in the prompt, and updates receive the failed checks without graph-based localization or update scoping.
    
    \item \textbf{Text-State Memory (TSM)} records generated files, failed checks, and update history as natural-language entries that are included in later generation and update prompts. It preserves project information but not explicit typed relations among states.
    
    \item \textbf{JSON-State} stores the same nodes, attributes, and typed relations as the Project State Graph in a JSON document. The complete JSON state is provided to the model, which decides which states are relevant and which files to revise without programmatic subgraph extraction, localization, or scoped update.
\end{itemize}

These variants separate three factors.
TSM tests whether retaining project history is sufficient.
JSON-State further tests whether storing the same structured states and typed relations is sufficient.
The comparison between JSON-State and \approach isolates the effect of graph operations, including task-scoped extraction, lookup-based localization, and bounded update scoping.
Table~\ref{tab:rq2_psg_ablation} reports the benchmark scores, and Table~\ref{tab:rq2_drift} reports the drift metrics.

\begin{table*}[!t]
\centering
\caption{Effect of the project state representation on Vision2Web. \textsc{TSM} denotes Text-State Memory and \textsc{JSON} denotes JSON-State.}
\label{tab:rq2_psg_ablation}
\setlength{\tabcolsep}{4pt}
\renewcommand{\arraystretch}{1.08}

\definecolor{DeepBlue}{RGB}{77,88,234}
\definecolor{LightBlue}{RGB}{235,241,250}

\resizebox{0.82\textwidth}{!}{
\begin{tabular}{l|ccccccccccccc}
\toprule
\multirow{2}{*}{\textbf{Method}} 
& \multicolumn{5}{c}{\textbf{Static Webpage}} 
& \multicolumn{4}{c}{\textbf{Interactive Frontend}} 
& \multicolumn{4}{c}{\textbf{Full-Stack Website}} \\
\cmidrule(lr){2-6} \cmidrule(lr){7-10} \cmidrule(lr){11-14}
& \textbf{Desktop} 
& \textbf{Tablet} 
& \textbf{Mobile} 
& \textbf{Avg} 
& \textbf{DSR}
& \textbf{VS} 
& \textbf{FS} 
& \textbf{Avg} 
& \textbf{DSR}
& \textbf{VS} 
& \textbf{FS} 
& \textbf{Avg} 
& \textbf{DSR} \\
\midrule

\multicolumn{14}{c}{\textbf{\textit{Doubao-Seed-1.8}}} \\
\midrule
w/o PSG & 61.4  & 57.7  & 51.0  & 56.7  & 100\%  & 28.3  & 28.7  & 28.5  & 97\%  & 17.7  & 10.2  & 14.0  & 100\%  \\
\textsc{TSM}     & 61.2  & 59.3  & 50.1  & 56.9  & 100\%  & 22.1  & 22.6  & 22.4  & 97\%  & 16.2  & 7.9   & 12.1  & 100\%  \\
\textsc{JSON}  & 62.4 & 58.1 & 53.9 & 58.1 & 100\% & 39.8 & 47.0 & 43.4 & 98.5\% & 27.1 & 28.7 & 27.9 & 96.3\% \\
\rowcolor{LightBlue}
\textcolor{DeepBlue}{\textbf{\approach}} 
& 63.7  & 57.5  & 58.1  & 59.8  & 100\%  & 50.6  & 55.7  & 53.2  & 98.5\%  & 35.6  & 40.3  & 38.0  & 100\% \\
\midrule

\multicolumn{14}{c}{\textbf{\textit{Gemini-3-Flash-Preview}}} \\
\midrule
w/o PSG & 59.1  & 55.0  & 45.6  & 53.2  & 100\%  & 38.6  & 36.7  & 37.7  & 100\%   & 36.7  & 21.0  & 28.9  & 96.3\%  \\
\textsc{TSM}     & 60.7  & 58.7  & 49.1  & 56.2  & 100\%  & 34.3  & 31.4  & 32.9  & 98.5\%  & 30.0  & 17.9  & 24.0  & 96.3\%  \\
\textsc{JSON}  & 62.1 & 62.7 & 63.6 & 62.8 & 100\% & 48.5 & 60.7 & 54.6 & 98.5\% & 44.1 & 32.5 & 38.3 & 96.3\% \\
\rowcolor{LightBlue}
\textcolor{DeepBlue}{\textbf{\approach}} 
& 65.0  & 66.4  & 68.9  & 66.8  & 100\%  & 53.4  & 72.2  & 62.8   & 100\%  & 51.0  & 40.8  & 45.9  & 96.3\%  \\
\midrule

\multicolumn{14}{c}{\textbf{\textit{Gemini-3.1-Pro-Preview}}} \\
\midrule
w/o PSG & 43.1  & 40.4  & 36.3  & 39.9  & 100\%  & 24.9  & 24.8  & 24.9  & 100\%   & 19.2  & 9.5   & 14.4  & 96.3\%  \\
\textsc{TSM}     & 47.7  & 44.2  & 40.1  & 44.0  & 100\%  & 26.8  & 28.3  & 27.6  & 98.5\%  & 17.0  & 9.6   & 13.3  & 96.3\%  \\
\textsc{JSON}  & 57.6 & 55.2 & 54.4 & 55.7 & 100\% & 43.4 & 48.0 & 45.7 & 100\% & 38.9 & 28.1 & 33.5 & 100\% \\
\rowcolor{LightBlue}
\textcolor{DeepBlue}{\textbf{\approach}} 
& 62.8  & 60.2  & 60.3  & 61.1  & 100\%  & 56.9  & 57.5  & 57.2  & 100\%  & 50.3  & 36.5  & 43.4  & 100\%  \\
\midrule

\multicolumn{14}{c}{\textbf{\textit{GPT-5.1}}} \\
\midrule
w/o PSG & 45.0  & 42.7  & 40.3  & 42.7  & 100\%  & 55.1  & 76.9  & 66.0  & 98.5\%  & 32.0  & 21.3  & 26.7  & 100\%  \\
\textsc{TSM}     & 46.2  & 42.7  & 41.0  & 43.3  & 100\%  & 50.0  & 68.7  & 59.4  & 100\%   & 27.2  & 17.7  & 22.5  & 96.3\%  \\
\textsc{JSON}  & 55.9 & 54.7 & 51.4 & 54.0 & 100\% & 55.6 & 80.0 & 67.8 & 100\% & 42.7 & 40.5 & 41.6 & 96.3\% \\
\rowcolor{LightBlue}
\textcolor{DeepBlue}{\textbf{\approach}} 
& 61.2  & 63.5  & 60.4  & 61.7  & 100\%  & 57.6  & 80.9  & 69.3  & 100\%  & 50.1  & 50.8  & 50.4  & 100\%  \\
\midrule

\multicolumn{14}{c}{\textbf{\textit{Qwen3.5-Plus}}} \\
\midrule
w/o PSG & 59.0  & 55.3  & 53.4  & 55.9  & 100\%  & 31.0  & 33.7  & 32.4  & 97\%  & 19.3  & 11.4  & 15.4  & 100\%  \\
\textsc{TSM}     & 59.8  & 56.4  & 55.3  & 57.2  & 100\%  & 25.1  & 26.8  & 26.0  & 97\%  & 21.5  & 12.3  & 16.9  & 100\%  \\
\textsc{JSON}  & 62.0 & 61.7 & 57.0 & 60.2 & 100\% & 44.5 & 53.3 & 48.9 & 97\% & 32.6 & 30.2 & 31.4 & 100\% \\
\rowcolor{LightBlue}
\textcolor{DeepBlue}{\textbf{\approach}} 
& 63.5  & 64.5  & 57.7  & 61.9  & 100\%  & 54.8  & 67.9  & 61.4  & 97\%  & 39.8  & 42.5  & 41.1  & 100\%  \\

\bottomrule
\end{tabular}
}

\vspace{-10px}
\end{table*}
\begin{table}[!t]
\centering
% [Original caption: Drift metrics on Vision2Web, averaged over the five foundation models (per-model results in our repository). CBR and TRR are computed on the final projects of all methods; VSDR and RR require explicit update rounds and are not defined for Claude Code and the +CC pipelines. $n_{\mathrm{pass}}$ is the number of passing test cases before an update, i.e., the denominator of VSDR.]
\caption{Drift metrics on Vision2Web, averaged over the five foundation models (per-model results in our repository). CBR and TRR are computed on the final projects of all methods; VSDR and RR require explicit update rounds and are not defined for Claude Code and the +CC pipelines. $n_{\mathrm{pass}}$ is the number of passing test cases before an update, i.e., the denominator of VSDR; because the update loops of the variants accept different numbers of patches, $n_{\mathrm{pass}}$ differs across rows, and VSDR should be read against it rather than in isolation.}
\label{tab:rq2_drift}
\definecolor{DeepBlue}{RGB}{77,88,234}
\definecolor{LightBlue}{RGB}{235,241,250}
\footnotesize
\setlength{\tabcolsep}{3pt}
\renewcommand{\arraystretch}{1.08}
\resizebox{0.75\linewidth}{!}{
\begin{tabular}{l|ccccc|ccccc}
\toprule
\multirow{2}{*}{\textbf{Method}}
& \multicolumn{5}{c|}{\textbf{Interactive Frontend}}
& \multicolumn{5}{c}{\textbf{Full-Stack Website}} \\
& \textbf{CBR}$\downarrow$ & \textbf{TRR} & \textbf{VSDR}$\downarrow$ & \textbf{RR}$\uparrow$ & $n_{\mathrm{pass}}$
& \textbf{CBR}$\downarrow$ & \textbf{TRR} & \textbf{VSDR}$\downarrow$ & \textbf{RR}$\uparrow$ & $n_{\mathrm{pass}}$ \\
\midrule
Best +CC pipeline
  & 46.6\% & 70.6\% & -- & -- & --
  & 65.1\% & 50.1\% & -- & -- & -- \\

Claude Code
  & 53.6\% & 58.7\% & -- & -- & --
  & 69.3\% & 30.6\% & -- & -- & -- \\

\midrule
w/o PSG
  & 50.0\% & 80.4\% & 5.7\% & 21.7\% & 516
  & 78.8\% & 69.1\% & 7.8\% & 16.7\% & 473 \\

\textsc{TSM}
  & 55.0\% & 79.1\% & 4.6\% & 22.9\% & 639
  & 80.8\% & 68.0\% & 6.9\% & 17.8\% & 677 \\

\textsc{JSON}
  & 31.4\% & 84.3\% & 2.4\% & 24.5\% & 347
  & 56.9\% & 74.2\% & 3.5\% & 19.1\% & 318 \\

\rowcolor{LightBlue}
\textcolor{DeepBlue}{\approach}
  & 22.2\% & 85.9\% & 0.9\% & 26.0\% & 643
  & 44.4\% & 75.8\% & 1.3\% & 20.6\% & 738 \\
\bottomrule
\end{tabular}
}
\vspace{-10px}
\end{table}

\noindent\textbf{Benchmark scores.}
\approach achieves the best overall performance across the three task levels.
Averaged over the five foundation models, it improves Avg over w/o PSG by 12.6 points on Static Webpage, 22.9 points on Interactive Frontend, and 23.9 points on Full-Stack Website tasks.
Compared with TSM, the corresponding gains are 10.7, 27.1, and 26.0 points.
On Static Webpage tasks, where a single page makes the generation input of w/o PSG nearly identical to that of \approach, w/o PSG scores below \approach without any update (w/o Update Loop in Table~\ref{tab:rq3_agent_ablation}) under four of the five models, e.g., 39.9 versus 60.8 with Gemini-3.1-Pro-Preview, indicating that unscoped updates degrade pages that were correct after generation.
Compared with JSON-State, \approach achieves higher Avg scores across the three task levels.
Because JSON-State contains the same structured information as the Project State Graph, its lower scores cannot come from the stored content.
The graph operations therefore matter in their own right, since they determine which states each generation step receives and which files an update may modify.

\noindent\textbf{Drift.}
The same pattern appears in the drift metrics.
Compared with the three variants, \approach achieves lower CBR and VSDR, and also reaches higher TRR and RR than all three.
The absolute counts behind VSDR are small, and Section~\ref{sec:threats} reads the rates against those counts.
Two features of the protocol shape how the CBR comparison should be read.
CBR is defined only over reachable triggers, so the higher TRR of \approach puts more test cases into the CBR denominator rather than fewer, and the lowest CBR is achieved on the hardest denominator.
CBR also remains driven mainly by generation, because verification observes rendered pages and never executes a workflow.
Storing project history in natural language does not help: \textsc{TSM} has a higher CBR than removing the project state entirely (55.0\% versus 50.0\% on Interactive Frontend tasks, at nearly identical TRR).

\begin{tcolorbox}[
  enhanced,
  colback=gray!15,
  colframe=gray!15,
  boxrule=0pt,
  arc=0pt,
  left=6pt,
  right=6pt,
  top=4pt,
  bottom=4pt,
  borderline west={2pt}{0pt}{gray!70}
]
\textbf{Answer to RQ2:}
The Project State Graph outperforms all three alternative representations, including JSON-State with the same information and the same budget, so the gain comes from the graph operations rather than from storing structured state.
\end{tcolorbox}

\subsection{RQ3: Operation Ablation}
\label{sec:rq3}

To answer RQ3, we examine the two stages that construct and use the Project State Graph.
We compare the full version of \approach with two variants:
\begin{itemize}
    \item \textbf{w/o Prototype Grounding} removes the grounding stage and generates the project directly from the visual prototypes and textual requirements. Since grounding jointly extracts page and interaction states, aligns them with prototype elements, and constructs the initial graph, this variant measures their combined contribution.
    
    \item \textbf{w/o Update Loop} stops after graph-guided generation and performs no verification or update. This variant therefore measures the combined contribution of render-evidence feedback and the subsequent scoped update.
\end{itemize}
All variants use the same foundation models, visual prototypes, textual requirements, and evaluation scripts.
Table~\ref{tab:rq3_agent_ablation} reports the detailed results.

\begin{table*}[!t]
\centering
\caption{Ablation study of key operations in \approach on Vision2Web.}
\label{tab:rq3_agent_ablation}
\setlength{\tabcolsep}{4pt}
\renewcommand{\arraystretch}{1.08}

\definecolor{DeepBlue}{RGB}{77,88,234}
\definecolor{LightBlue}{RGB}{235,241,250}

\resizebox{0.83\textwidth}{!}{
% [Column rule added after the method column to match Tables \ref{tab:rq1_main_results} and \ref{tab:rq2_psg_ablation}. Original: \begin{tabular}{lccccccccccccc}]
\begin{tabular}{l|ccccccccccccc}
\toprule
\multirow{2}{*}{\textbf{Method}}
& \multicolumn{5}{c}{\textbf{Static Webpage}}
& \multicolumn{4}{c}{\textbf{Interactive Frontend}}
& \multicolumn{4}{c}{\textbf{Full-Stack Website}} \\
\cmidrule(lr){2-6} \cmidrule(lr){7-10} \cmidrule(lr){11-14}
& \textbf{Desktop}
& \textbf{Tablet}
& \textbf{Mobile}
& \textbf{Avg}
& \textbf{DSR}
& \textbf{VS}
& \textbf{FS}
& \textbf{Avg}
& \textbf{DSR}
& \textbf{VS}
& \textbf{FS}
& \textbf{Avg}
& \textbf{DSR} \\
\midrule

\multicolumn{14}{c}{\textbf{\textit{Doubao-Seed-1.8}}} \\
\midrule
w/o Prototype Grounding & 59.4 & 57.3 & 50.8 & 55.8 & 99.0\% & 20.4 & 9.8  & 15.1 & 98.5\% & 15.0 & 8.2  & 11.6 & 100\% \\
w/o Update Loop   & 59.6 & 54.8 & 49.4 & 54.6 & 100\%  & 47.1 & 54.4 & 50.8 & 98.5\% & 30.4 & 26.4 & 28.4 & 100\% \\
\rowcolor{LightBlue}
\textcolor{DeepBlue}{\textbf{\approach}} & 63.7 & 57.5 & 58.1 & 59.8 & 100\% & 50.6 & 55.7 & 53.2 & 98.5\% & 35.6 & 40.3 & 38.0 & 100\% \\
\midrule

\multicolumn{14}{c}{\textbf{\textit{Gemini-3-Flash-Preview}}} \\
\midrule
w/o Prototype Grounding & 56.2 & 51.6 & 48.0 & 51.9 & 100\% & 16.4 & 13.7 & 15.1 & 98.5\% & 18.6 & 8.4  & 13.5 & 96.3\% \\
w/o Update Loop   & 63.4 & 58.5 & 62.7 & 61.5 & 100\% & 51.7 & 59.5 & 55.6 & 100\%  & 41.2 & 36.7 & 38.9 & 96.3\% \\
\rowcolor{LightBlue}
\textcolor{DeepBlue}{\textbf{\approach}} & 65.0 & 66.4 & 68.9 & 66.8 & 100\% & 53.4 & 72.2 & 62.8 & 100\% & 51.0 & 40.8 & 45.9 & 96.3\% \\
\midrule

\multicolumn{14}{c}{\textbf{\textit{Gemini-3.1-Pro-Preview}}} \\
\midrule
w/o Prototype Grounding & 56.1 & 53.7 & 48.4 & 52.7 & 100\% & 38.2 & 36.0 & 37.1 & 98.5\% & 34.3 & 15.2 & 24.8 & 100\% \\
w/o Update Loop   & 63.4 & 58.5 & 60.6 & 60.8 & 100\% & 52.3 & 53.1 & 52.7 & 100\%  & 44.2 & 30.9 & 37.6 & 100\% \\
\rowcolor{LightBlue}
\textcolor{DeepBlue}{\textbf{\approach}} & 62.8 & 60.2 & 60.3 & 61.1 & 100\% & 56.9 & 57.5 & 57.2 & 100\% & 50.3 & 36.5 & 43.4 & 100\% \\
\midrule

\multicolumn{14}{c}{\textbf{\textit{GPT-5.1}}} \\
\midrule
w/o Prototype Grounding & 56.0 & 53.1 & 52.6 & 53.9 & 100\% & 38.1 & 40.3 & 39.2 & 100\% & 34.7 & 15.5 & 25.1 & 96.3\% \\
w/o Update Loop   & 59.0 & 54.6 & 54.1 & 55.9 & 100\% & 53.5 & 78.9 & 66.2 & 100\% & 47.8 & 49.1 & 48.5 & 100\% \\
\rowcolor{LightBlue}
\textcolor{DeepBlue}{\textbf{\approach}} & 61.2 & 63.5 & 60.4 & 61.7 & 100\% & 57.6 & 80.9 & 69.3 & 100\% & 50.1 & 50.8 & 50.4 & 100\% \\
\midrule

\multicolumn{14}{c}{\textbf{\textit{Qwen3.5-Plus}}} \\
\midrule
w/o Prototype Grounding & 59.6 & 56.5 & 53.2 & 56.4 & 100\% & 25.5 & 19.6 & 22.6 & 97.0\% & 31.4 & 13.0 & 22.2 & 96.3\% \\
w/o Update Loop   & 63.4 & 56.9 & 54.1 & 58.1 & 100\% & 49.9 & 53.5 & 51.7 & 97.0\% & 37.9 & 37.2 & 37.6 & 100\% \\
\rowcolor{LightBlue}
\textcolor{DeepBlue}{\textbf{\approach}} & 63.5 & 64.5 & 57.7 & 61.9 & 100\% & 54.8 & 67.9 & 61.4 & 97.0\% & 39.8 & 42.5 & 41.1 & 100\% \\

\bottomrule
\end{tabular}
}
\end{table*}

Table~\ref{tab:rq3_agent_ablation} shows that both stages contribute to \approach.
Averaged over the five foundation models, removing prototype grounding decreases Avg from 60.8 to 25.8 on Interactive Frontend tasks and from 43.8 to 19.4 on Full-Stack Website tasks, whereas removing verification and scoped update decreases it to 55.4 and 38.2, respectively.
Without grounding, generation has no interaction states linked to prototype elements and routes, and on Interactive Frontend tasks FS drops more than VS under every model (e.g., from 55.7 to 9.8 with Doubao-Seed-1.8).
Most of the gain therefore comes from constructing the project state before generation, while the update stage corrects the defects that remain.
Removing the update loop, including the visual judge, lowers VS by only 3.8 and 5.1 points on average on Interactive Frontend and Full-Stack Website tasks, so most of the VS advantage over the baselines is obtained before any judge feedback.

\begin{tcolorbox}[
  enhanced,
  colback=gray!15,
  colframe=gray!15,
  boxrule=0pt,
  arc=0pt,
  left=6pt,
  right=6pt,
  top=4pt,
  bottom=4pt,
  borderline west={2pt}{0pt}{gray!70}
]
\textbf{Answer to RQ3:}
Both stages contribute, and prototype grounding dominates: it accounts for far more of the Avg gain than verification and scoped update.
\end{tcolorbox}

\subsection{RQ4: Correctness of the Project State Graph}
\label{sec:rq4}

RQ1--RQ3 evaluate the Project State Graph through the websites it helps generate.
However, later generation and update may compensate for errors in the graph, so end-to-end performance does not establish that the graph itself is correct.
We therefore directly evaluate the nodes and relations in the initial graph $\mathcal{G}_0$ produced by prototype grounding.

\noindent\textbf{Sample.}
We manually inspect the initial graphs of 50 tasks, including all 27 Full-Stack Website tasks and 23 Interactive Frontend tasks sampled uniformly at random.
We exclude Static Webpage tasks because they contain no interaction states.
For each selected task, we annotate one model run and balance the 50 runs across the five foundation models, with 10 runs per model.

\noindent\textbf{Annotation procedure.}
Two authors independently construct a reference graph for each task using only the visual prototypes, textual requirements, and the benchmark's expert-written functional workflows.
The generated graph is not shown while constructing the reference, allowing the annotators to identify both incorrect states and states or relations that the generated graph misses.
After completing the references independently, the annotators compare their annotations and resolve disagreements through discussion to obtain the final gold graph.
The annotation guideline and task-level annotations are included in the replication package.
For page and region states, annotators identify the pages and visually distinguishable functional regions required by the prototypes and requirements.
For text and asset states, they record visible text and required asset slots, such as product images, icons, and banners.
For interaction states, the reference is derived from the benchmark's expert-written workflows, which specify the triggering element, action sequence, and expected target page or state.
These workflows are used only for annotation and are never exposed to \approach during generation.

We evaluate three semantic relation types.
A \texttt{grounds\_to} or \texttt{triggers} edge is correct when an interaction state is linked to the specific prototype element that initiates the corresponding workflow.
Linking an interaction only to its source page is not counted as correct when the triggering element can be identified.
A \texttt{depends\_on} edge for an asset state is correct when the linked resource provides the content required by that asset slot.
A \texttt{depends\_on} edge for an interaction state is correct when the linked route serves the target page or state specified by the requirement and the corresponding workflow.

\noindent\textbf{Metrics.}
Generated and gold nodes are matched one-to-one by semantic identity rather than exact wording.
Page and region nodes are matched by page identity and region function, text and asset nodes by visible content or required slot, and interaction nodes by the corresponding workflow.
A matched generated node is a true positive, an unmatched generated node is a false positive, and an unmatched gold node is a false negative.
For edges, a generated edge is a true positive only when its source node, target node, and relation type match a gold relation.
Unsupported generated edges are false positives, while gold relations absent from the generated graph are false negatives.
We report precision, recall, and F1 for each node and edge type, together with the overall edge-level F1.

To measure inter-annotator agreement before discussion, we canonicalize the nodes and relations identified by either annotator, take their union, and compare the two annotators' independent present/absent judgments for each candidate.
The resulting Cohen's $\kappa$ is 92\%.

\begin{table*}[t]
\centering
\caption{Correctness of the initial Project State Graph on 50 manually annotated tasks.}
\label{tab:rq4_psg_correctness}
\footnotesize

\begin{minipage}[t]{0.38\textwidth}
\centering
\textbf{(a) Node correctness}

\vspace{2pt}
\setlength{\tabcolsep}{4pt}
\begin{tabular}{lrrrr}
\toprule
\textbf{Node Type} & \textbf{\#Gold} & \textbf{P} & \textbf{R} & \textbf{F1} \\
\midrule
Page/region  & 3468 & 99.0\% & 98.4\% & 98.7\% \\
Text/asset   & 9350 & 98.1\% & 96.4\% & 97.2\% \\
Interaction  & 1138 & 97.2\% & 95.9\% & 96.5\% \\
\bottomrule
\end{tabular}
\end{minipage}
\hfill
\begin{minipage}[t]{0.58\textwidth}
\centering
\textbf{(b) Edge correctness}

\vspace{2pt}
\setlength{\tabcolsep}{4pt}
\begin{tabular}{lrrrr}
\toprule
\textbf{Edge Type} & \textbf{\#Gold} & \textbf{P} & \textbf{R} & \textbf{F1} \\
\midrule
\texttt{grounds\_to}/\texttt{triggers}
    & 1829 & 97.2\% & 94.7\% & 95.9\% \\
\texttt{depends\_on} (asset)
    & 1267 & 98.0\% & 96.6\% & 97.3\% \\
\texttt{depends\_on} (interaction)
    & 1024 & 94.6\% & 91.6\% & 93.1\% \\
\midrule
\textbf{All edges}
    & 4120 & 96.8\% & 94.5\% & 95.6\% \\
\bottomrule
\end{tabular}
\end{minipage}

\vspace{-10px}
\end{table*}

\noindent\textbf{Results.}
Table~\ref{tab:rq4_psg_correctness} shows that the initial Project State Graph achieves an overall edge-level F1 of 95.6\%.
The per-type results identify which relations are constructed reliably and where the remaining errors are concentrated.
We further inspect the false positives and false negatives to distinguish errors caused by missing states, incorrect prototype-element grounding, and incorrect dependency relations.
% [Original: To trace how construction errors propagate, we group annotated workflows by the correctness of their \texttt{grounds\_to}/\texttt{triggers} edges: workflows with a correct edge pass 56.2\% of their test cases, compared with 4.7\% for those with an incorrect or missing edge.
% This comparison is associational rather than controlled: a workflow whose trigger element can be identified unambiguously tends to be a simpler workflow, so part of the gap may reflect workflow difficulty rather than the edge itself.
% The gap is large enough that difficulty alone is unlikely to account for all of it, but we do not isolate the two factors here.]
% [TODO(authors): report the number of workflows in each group, i.e., those whose grounds_to/triggers edge is correct and those whose edge is incorrect or missing, so that the pass rates below are not read without their support.]
To trace how construction errors propagate, we group annotated workflows by the correctness of their \texttt{grounds\_to}/\texttt{triggers} edges: workflows with a correct edge pass 56.2\% of their test cases, compared with 4.7\% for those with an incorrect or missing edge.
The comparison is associational rather than controlled, since ambiguously grounded workflows may also be the harder ones, but the gap is large enough that difficulty alone is unlikely to explain it.

\begin{tcolorbox}[
  enhanced,
  colback=gray!15,
  colframe=gray!15,
  boxrule=0pt,
  arc=0pt,
  left=6pt,
  right=6pt,
  top=4pt,
  bottom=4pt,
  borderline west={2pt}{0pt}{gray!70}
]
\textbf{Answer to RQ4:} On 50 annotated tasks, the initial Project State Graph reaches an edge-level F1 of 95.6\%, while depends\_on(interaction) is the weakest relation type, with an F1 of 93.1\%.
\end{tcolorbox}

\section{Discussion}
\label{sec:discussion}

\subsection{Dependence on the Foundation Model}
\label{sec:disc_model}

The absolute performance of \approach varies with the backbone, e.g., Full-Stack Website Avg from 38.0 to 50.4, so we examine how the benefit of the graph, measured as \approach minus w/o PSG, depends on the model.
On Full-Stack Website tasks, the gain is 17.0 to 29.0 points across the five models and tends to be smaller for backbones with a higher w/o PSG score, but every model gains at least 17 points, so the graph does not only compensate for weak models.
On Interactive Frontend tasks, the gain is 24.7 to 32.3 points for four models but only 3.3 for GPT-5.1, whose w/o PSG variant already reaches 66.0, while the same model gains 23.7 points on Full-Stack Website tasks.
Strong models may track these correspondences implicitly in small projects, or the smaller task set may leave less headroom for them; the data cannot separate the two.
We observe the same trend at the task level, where the gain over w/o PSG, averaged over the five models, is positively correlated with the number of workflows ($\rho=0.62$, $n=93$, $p<0.001$) and pages ($\rho=0.47$, $n=93$, $p<0.001$) per task.
Among the components, grounding contributes most on Interactive Frontend and Full-Stack Website tasks (18.6 to 47.7 points), the update loop least (1.9 to 9.7), and the Project State Graph as a whole varies most on small projects (3.3 to 32.3 points of \approach minus w/o PSG on Interactive Frontend tasks).
On Static Webpage tasks both components stay low, as expected without interaction state.

\subsection{Evaluation on WebGen-Bench}
\label{sec:disc_webgen}

WebGen-Bench~\cite{lu2025webgenbench} contains 101 website-generation instructions in 13 categories with 647 agent-executed test cases, but no visual prototypes.
We therefore run \approach without prototype states: interaction and page states come from the task instruction, and pixel comparison and the visual judge are disabled.
We compare \approach with Claude Code, the only baseline that also runs without prototypes.
Run once, \approach reaches a test-case accuracy of 49.0\% against 40.1\% for Claude Code.
The gain is smaller than on Vision2Web, as expected without prototype grounding, but it indicates that a structured project state remains useful when the task is specified only in natural language.

\subsection{Cost and Scalability}
\label{sec:disc_cost}

For a project with $p$ pages, the number of LLM calls grows linearly with the number of pages and update rounds. Grounding uses up to six calls per page and four project-level calls, generation uses three calls per page plus one call for each shared file, and each update round invokes the visual judge once per page with one or two additional calls for each issue group.
Table~\ref{tab:cost} reports the measured cost.
Against Claude Code, \approach uses $1.7\times$ the tokens and $3.4\times$ the runtime on Full-Stack Website tasks, with an average API cost of \$20.93 per task.
The more informative comparison is with w/o PSG, which shares the same generation and update budget: \approach costs about $1.03\times$ as much on both tokens and runtime, so the graph is nearly free relative to the variant it replaces.

\begin{table}[!t]
\centering
\caption{Average cost per task over the five foundation models. Token counts are in thousands; costs use the per-model list prices on the access dates in the replication package. \emph{Graph ops} is the non-LLM runtime of graph construction, localization, and scope computation. The Avg column repeats Table~\ref{tab:rq1_main_results} and is not a cost measure.}
\label{tab:cost}
\footnotesize
\setlength{\tabcolsep}{3pt}
\resizebox{\linewidth}{!}{
\begin{tabular}{ll|rrrrrr|r}
\toprule
\textbf{Level} & \textbf{Method} & \textbf{Calls} & \textbf{Input (K)} & \textbf{Output (K)} & \textbf{Cost (\$)} & \textbf{Time (min)} & \textbf{Graph ops (ms)} & \textbf{Avg (Tab.~\ref{tab:rq1_main_results})} \\
\midrule
\multirow{4}{*}{Interactive Frontend}
 & Best +CC pipeline & 70.3 & 3492.8 & 38.6 & 9.50 & 19.13 & -- & 34.6 \\
 & Claude Code       & 58.4 & 2847.6 & 29.7 & 7.71 & 14.78 & -- & 20.1 \\
 & w/o PSG           & 77.8 & 4536.7 & 89.6 & 13.13 & 49.27 & -- & 37.9 \\
 & \approach         & 78.9 & 4648.3 & 92.2 & 13.46 & 50.82 & 673.4 & 60.8 \\
\midrule
\multirow{4}{*}{Full-Stack Website}
 & Best +CC pipeline & 101.5 & 5294.7 & 58.4 & 14.40 & 29.63 & -- & 21.2 \\
 & Claude Code       & 82.7 & 4346.9 & 45.6 & 11.78 & 22.87 & -- & 6.8 \\
 & w/o PSG           & 110.9 & 7041.8 & 136.6 & 20.34 & 76.36 & -- & 19.9 \\
 & \approach         & 112.6 & 7247.3 & 140.8 & 20.93 & 78.27 & 1187.6 & 43.8 \\
\bottomrule
\end{tabular}
}
\vspace{-10px}
\end{table}

The graph operations are lookups and a two-hop traversal, linear in graph size, and add little runtime (Table~\ref{tab:cost}). The bottleneck is the shared files: page files receive page-scoped context, whereas the theme, interaction script, and server are generated from all pages and interactions, so their context grows with project size. Requirement documents with hundreds of scenarios~\cite{kong2026arc} will therefore require finer-grained scoping for shared files.

\section{Threats to Validity}
\label{sec:threats}

\textbf{Internal validity.}
The main threat is that each configuration is generated and evaluated once, so run-to-run variance is not quantified, the pooled differences of a few points in Section~\ref{sec:rq3} and the drift metrics of Section~\ref{sec:rq2} should be read as point estimates, and no comparison in this paper is a tested difference.
The drift metrics are affected most, because VSDR is small by construction: on Interactive Frontend tasks the \approach and JSON-State rows correspond to roughly 6 and 8 regressed test cases out of 643 and 347 passing cases, so a small change in evaluator behavior could move the rates appreciably even if the ranking is stable in all five models.
We did not measure the evaluator's own flip rate on a fixed project, which would give the noise floor against which VSDR should be judged.
Static UI-to-code methods are evaluated as +CC pipelines on interactive tasks, so those results characterize the pipelines, and we report the static stage separately.
Claude Code runs with non-Anthropic models through a proxy, whose format translation may lower its performance; we follow the Vision2Web configuration and release all transcripts.
RQ4 relies on human judgments, which may introduce subjective variation; we mitigate this through independent annotation and report inter-annotator agreement, although the gold graphs are built by the authors, who know the schema of the Project State Graph and may align their references with its ontology more readily than an external annotator would.
The RQ3 variants remove extraction and alignment, and feedback and scoping, together; JSON-State isolates the graph operations as a whole, but no variant removes write protection alone.
The propagation comparison in Section~\ref{sec:rq4} is associational, since incorrectly grounded workflows may also be the harder ones.

\noindent\textbf{Construct validity.}
We follow the official Vision2Web protocol; VS relies on a VLM judge, and FS measures workflow-level correctness through expert-written test cases, which do not cover every implied behavior or qualities such as maintainability and accessibility.
The drift metrics capture drift only where a test case exercises the affected correspondence, and VSDR is undefined for methods without update rounds, for which we report CBR.
CBR counts every failure after a reachable trigger, including backend faults, and test cases of a workflow share one browser session, so an earlier failure can affect later cases.
The internal checks of \approach do not execute workflows, so behavioral defects are addressed only by the links established during generation.

\noindent\textbf{External validity.}
Vision2Web tasks contain up to 7 pages and WebGen-Bench has no prototypes, so scalability to larger projects is not evaluated (Section~\ref{sec:disc_cost}).
\approach generates plain HTML, CSS, and JavaScript with an optional Node.js server and is not evaluated on frameworks such as React or Vue.
Model APIs change over time, so we report identifiers and access dates; the websites underlying Vision2Web may also appear in training data, which would affect all methods alike.

\section{Related Work}
\label{sec:related}

\textbf{Deployable Website Generation.}
Deployable website generation turns visual prototypes and textual requirements into an executable website project, and existing approaches fall into UI-to-code methods and agent-based methods.
UI-to-code research ranges from reverse engineering mobile interfaces~\cite{nguyen2015reverse,moran2018machine} to multimodal LLMs evaluated on Design2Code~\cite{si2025design2code}, Web2Code~\cite{yun2024web2code}, and WebCode2M~\cite{gui2025webcode2m}.
Static methods reconstruct layouts: UICopilot~\cite{gui2025uicopilot} generates code hierarchically, DCGen~\cite{wan2025divide} and LayoutCoder~\cite{wu2025mllm} decompose screenshots or follow layout information, and UIOrchestra~\cite{yue2025uiorchestra} and ScreenCoder~\cite{jiang2025screencoder} coordinate specialized agents.
Interactive methods extend the output to user actions: Interaction2Code~\cite{xiao2025interaction2code} and DesignBench~\cite{xiao2025designbench} evaluate interactive generation and front-end editing, WebVIA~\cite{xu2025webvia} validates interactions through exploration, and Vision2Web~\cite{he2026vision2web} covers multi-page frontends and full-stack websites.
These methods take a page or an interface as the unit of generation, so the correspondence between a prototype element and the files, routes, and workflows implementing it is not maintained across pages.
Agent-based methods include general coding agents, such as SWE-agent~\cite{yang2024swe}, OpenHands~\cite{wang2025openhands}, AutoCodeRover~\cite{zhang2024autocoderover}, Agentless~\cite{xia2025agentless}, and Claude Code~\cite{anthropic2026claudecode}, which edit code using tool feedback, program structure, and tests, and specification-to-repository frameworks.
ChatDev~\cite{qian2024chatdev} and MetaGPT~\cite{hong2024metagpt} pass design documents and code between role-playing agents, and ARC~\cite{kong2026arc} compiles multimodal requirement documents written in a domain-specific language into runnable systems through a bidirectional test-driven workflow with traceability across requirements, design, and code, scaling to hundreds of scenarios.
ARC and \approach are complementary: ARC organizes generation around DSL requirement scenarios and synthesized tests, whereas \approach organizes it around prototype elements and verifies rendered pages against the prototypes; because Vision2Web provides no DSL documents, we compare them conceptually.
In all agent-based methods, the project state is organized around code, tests, or requirement documents, so a mismatch between the rendered pages and the prototypes cannot be traced back to the prototype element and files it concerns.

\noindent\textbf{Traceability and Change Impact Analysis.}
Requirements traceability links requirements, design, code, and tests~\cite{cleland2014traceability}, usually by recovering the links after the artifacts exist with information retrieval~\cite{antoniol2002recovering}, deep learning~\cite{guo2017semantically}, or pre-trained language models~\cite{lin2021traceability}.
Change impact analysis estimates the entities a change affects through reachability over dependences~\cite{li2013survey} or execution traces~\cite{law2003whole}.
Fault localization takes the complementary direction, proceeding from a failure to its likely causes~\cite{wong2016survey}, and regression testing re-checks what a change may break~\cite{yoo2012regression}.
Repository-level LLM approaches bring such structure to coding: CodePlan~\cite{bairi2024codeplan} plans edits over a dependency graph with incremental impact analysis, and RepoGraph~\cite{ouyang2025repograph} provides a repository-level code graph.
For GUI and web software, consistency checkers detect design violations~\cite{moran2018automated} and cross-browser issues~\cite{choudhary2013x}, and search-based techniques repair layout issues~\cite{mahajan2017xfix}.
What these techniques share is that they recover links among textual artifacts after development, or start from a code change or a failing test.
\approach builds on them but differs in where the links come from and what starts the analysis: it fixes cross-modal trace links before code is written, seeds localization and a two-hop impact set with render evidence rather than with a code change, and write-protects files outside the impact set.

\section{Conclusion}

In this paper, we study deployable website generation from visual prototypes and textual requirements.
We identify \textit{prototype-to-project state drift} as a key challenge in this task, in which visual prototype elements, project files, interaction workflows, dependency relations, and render evidence lose their correspondence across generation, verification, and update, and we make it measurable through the correspondence break rate and the verified-state drift rate.
We propose \approach, which maintains these correspondences in a Project State Graph whose trace links are fixed before code is written and whose impact analysis starts from render evidence and confines each update to its scope.
On Vision2Web with five foundation models, \approach improves the overall score over the strongest baseline under each model by 26.1 and 22.6 points on Interactive Frontend and Full-Stack Website tasks, lowers the correspondence break rate from 44.2\% to 33.3\% and the verified-state drift rate from 3.0\% to 1.1\%, and builds graphs with an edge-level F1 of 95.6\% on an annotated sample.
The comparison with a JSON record holding the same information attributes the gains to the graph operations, and the gains are largest where projects span multiple pages and workflows.
Two extensions follow from our analysis: executing the extracted interaction states during verification, so that behavioral failures also become render evidence, and applying page-scoped context to shared files, so that \approach scales to larger projects.

% =========================================================
% References
% =========================================================

\bibliographystyle{ACM-Reference-Format}
\bibliography{references}

\end{document}